# Beneficial influence of Hf and Zr additions to Nb4at.%Ta on the vortex pinning of Nb$_3$Sn with and without an O source

Shreyas Balachandran[1*], Chiara Tarantini[1], Peter J. Lee[1], Fumitake Kametani[1,2], Yi-Feng Su[1], Benjamin Walker[1,2], William L. Starch[1], and David C. Larbalestier[1,2]

[1] National High Magnetic Field Laboratory, Tallahassee, FL 32310, USA.
[2] FAMU/FSU College of Engineering, Tallahassee, FL 32310, USA.

E-mail: shreyasb@asc.magnet.fsu.edu

**Abstract**

Here we show that addition of Hf to Nb4Ta can significantly improve the high field performance of Nb$_3$Sn, making it suitable for dipole magnets for a machine like the 100 TeV future circular collider (FCC). A big challenge of the FCC is that the desired non-Cu critical current density ($J_c$) target of 1500 A/mm$^2$ (16 T, 4.2 K) is substantially above the best present Nb$_3$Sn conductors doped with Ti or Ta (~1300 A/mm$^2$ in the very best sample of the very best commercial wire). Recent success with internal oxidation of Nb-Zr precursor has shown significant improvement in the layer $J_c$ of Nb$_3$Sn wires, albeit with the complication of providing an internal oxygen diffusion pathway and avoiding degradation of the irreversibility field $H_{\mathrm{Irr}}$. We here extend the Nb1Zr oxidation approach by comparing Zr and Hf additions to the standard Nb4Ta alloy of maximum $H_{c2}$ and $H_{\mathrm{irr}}$. Nb4Ta rods with 1Zr or 1Hf were made into monofilament wires with and without SnO$_2$ and their properties measured over the entire superconducting range at fields up to 31 T. We found that group IV alloying of Nb4Ta does raise $H_{\mathrm{Irr}}$, though adding O$_2$ still degrades this slightly. As noted in earlier Nb1Zr work with an O source, the pinning force density $F_p$ is strongly enhanced and its peak value shifted to higher field by internal oxidation. A surprising result of this work is that we found better properties in Nb4Ta1Hf without SnO$_2$, $F_{\mathrm{pMax}}$ achieving 2.35 Times that of the standard Nb4Ta alloy, while the oxidized Nb4Ta1Zr alloy achieved 1.54 times that of the Nb4Ta alloy. The highest layer $J_c$(16 T, 4.2 K) of 3700 A/mm$^2$ was found in the SnO$_2$-free wire made with Nb4Ta1Hf alloy. Using a standard A15 cross-section fraction of 60% for modern PIT and RRP wires, we estimated that a non-Cu $J_c$ of 2200 A/mm$^2$ is obtainable in modern conductors, well above the 1500 A/mm$^2$ FCC specification. Moreover, since the best properties were obtained without SnO$_2$, the Nb4Ta1Hf alloy appears to open a straightforward route to enhanced properties in Nb$_3$Sn wires manufactured by virtually all the presently used commercial routes employed today.

## 1. Introduction

The Future Circular collider (FCC) or upgrades to the Large Hadron Collider (LHC) require Nb$_3$Sn dipole magnets to operate up to 16 T [1-3], an operating range about twice the Nb-Ti capability of the present LHC magnets [4], and about 5 T higher than the Nb$_3$Sn magnets planned for the High Luminosity LHC upgrade [5,6]. To achieve such not yet achieved dipole fields requires magnet conductors with the not yet achieved high current density ($J_c$) of 1500 A/mm$^2$ at 16 T (4.2 K), while simultaneously maintaining high Cu stabilizer residual resistivity ratio (RRR) $\geq$ 100 and a large strand diameter of order 1 mm [7]. Recent rethinking of the heat treatment of commercial rod restack process (RRP) conductors has shown that a better optimized Sn mixing heat treatment can allow $J_c$ (16 T, 4.2 K) as high as 1300 A/mm$^2$ in specially selected 0.8 mm diameter wires, a significant 20% increase over the best prior standard heat treatment [8]. However, real conductor production has significant spread so there is little hope to achieve FCC specification by this route alone. Powder-In-Tube (PIT) conductor studies indicate generally similar properties to RRP® conductors but with slightly lower performance [9]. The present optimization focus of RRP and PIT commercial conductors is primarily to improve the quality and quantity of the Nb$_3$Sn in the reaction pack while maintaining high RRR. In summary, there is little expectation that present RRP or PIT conductors can achieve the very challenging 1500 A/mm$^2$ at 16 T, 4.2 K FCC specification.

Analysing the challenges emphasizes that high field performance of Nb$_3$Sn conductors requires minimal loss of upper critical field ($H_{c2}$) or the more practical irreversibility field ($H_{Irr}$). $H_{c2}$ sets the thermodynamic upper limit of superconductivity: pure binary Nb$_3$Sn has a $H_{c2}$ of 23.4 T (4.2 K), whereas doping with Group IV (Ti, Hf, Zr) or Group V (Mo, Ta) elements increases the $H_{c2}$ up to ~27 T [10,11]. Ta [12,13] and Ti [14] doping were adopted in PIT and RRP wires and also increase high-field $J_c$ [15]. However, stoichiometry variation across the A15 layer [16,17] and disorder induced by dopants both broaden the $H_{c2}$ distribution [18] and lower $H_{Irr}$ [11]. Nb$_3$Sn conductor optimization involves multiple parameters and understanding of averaged property distributions, reasons why property improvement has somewhat stagnated in recent years. Since there is little sign that $H_{c2}$ can be further enhanced beyond that possible with Ta and/or Ti [11,19], the main issue for the present is clearly to maximize the uniformity of the A15 layer and to avoid any degradation of $H_{Irr}$.

A separate issue is the value of significant enhancement of vortex pinning in the A15 layer. In the general absence of $J_c$ measurements in the 20-25 T range up to $H_{Irr}$, an important indication of enhanced vortex pinning is an upward shift of the field at which the pinning force density $F_p$ (= $|J_c \times \mu_0 H|$) is maximum ($H_{Max}$), typically 4-5 T. The history of attempts to enhance vortex pinning in Nb$_3$Sn is quite lengthy. A common theme has been to try to reduce the A15 grain size *and* to introduce additional pinning centres (APCs) into the Nb$_3$Sn. Ta [20] and Cu [21,22] inclusions were tried without great benefit to $J_c$, although major effects on enhancing $H_{Max}$ were reported by Rodrigues [22]. The most effective recent results are those observed by irradiation. Transport $J_c$ of 2000 A/mm$^2$ at 16 T was achieved in Ti-doped RRP strands after neutron irradiation [19], indicating that point pinning is a route to the FCC specification (though not in a practical conductor fabrication process). The most promising recent results are perhaps those obtained by insulating oxide nanoparticles created during the reaction heat treatment using a process first developed for a Nb-Zr alloy for a A15 Tape substrate in 1994 at GE [23,24] and also implemented more than a decade ago in experimental filamentary conductors [25-27]. In this route, Zr, originally in dilute solid solution in the Nb, is internally oxidized to insulating ZrO$_2$ point pins during the A15 reaction heat treatment. Oxygen can be introduced through an easily decomposable oxide such as SnO$_2$. Such particles increase the pinning force due to a combination of decreased A15 grain size below 50 nm and creation of insulating ZrO$_2$ point pins [28]. The encouraging results for high field performance are the shift of $H_{Max}$ to higher fields [28,29] and significantly enhanced A15 layer $J_c$ [28,30]. However, $H_{Irr}$ of conductors made with binary Nb-Zr alloy was suppressed with respect to clean Nb$_3$Sn and was far below that of modern Ta/Ti-doped Nb$_3$Sn [8,14]. Given this $H_{Irr}$ degradation, ternary alloys like Nb4Ta1Zr are an obvious next step.

This impasse for the FCC conductor motivated the principal questions of this work:
a) Is it possible to maintain the high $H_{Irr}$ of present commercial Nb$_3$Sn conductors made with Nb4Ta/Ti alloys while increasing the pinning site density with ternary alloys based on Nb4Ta-X (X being Zr or Hf)?
b) Does the introduction of oxygen in the system have detrimental effects on the irreversibility field?
c) To what extent are A15 grain size refinement and point pinning effects of insulating ZrO$_2$ or HfO$_2$ pins individually separable and additive?

In the work reported below we show that Hf is an even more powerful additive to Nb4Ta than Zr and that an external oxidant is not necessary to greatly enhance vortex pinning. We conclude that the FCC specification is attainable in present commercial Nb$_3$Sn conductor designs without internal oxygen sources.

**2. Experimental methods**

We prepared custom alloys with composition Nb4Ta1$X$, using Hf, not just Zr, because they both have high oxygen affinity and can create $X$O$_2$ precipitates in the presence of SnO$_2$. Our Nb$_3$Sn reaction package is a mixture of Cu-Sn powder surrounding the alloy rod inside a Ta/Cu tube shown in the inset of Figure 5. Monofilaments were made with and without SnO$_2$ mixed into the Cu-Sn powder. The molar volume ratio of SnO$_2$ was based on the calculation that 3% of the alloy rod would transform to A15 during heat treatment producing a reaction layer of about 3000-4000 μm². The monofilament wires were drawn down to a final diameter of 2 mm and underwent a two-step heat treatment at 550°C/100h and 670°C/100h. Vibrating-sample magnetometer (VSM) characterizations were performed up to 31 T at the National High Magnetic Field Laboratory to obtain the entire hysteresis loop from which the pinning force and irreversibility fields were calculated. The microstructures were evaluated in a Zeiss 1540 EsB scanning electron microscope (SEM) and a JEOL ARM200cF transmission electron microscope (TEM). Transverse wire cross-sections were prepared by metallographic polishing and then imaged with a solid-state Back-Scattered Electron (BSE) detector, followed by digital image analysis. The TEM specimens were prepared by mechanical polishing on diamond lapping films, followed by Ar ion milling. In the following we will refer to the samples according to the additions with respect to the binary compound (e.g. Ta-Hf is the wire made with Nb4Ta1Hf alloy, while Ta-Zr-$SnO_2$ refers to the wire prepared with Nb4Ta1Zr alloy and $SnO_2$ powder).

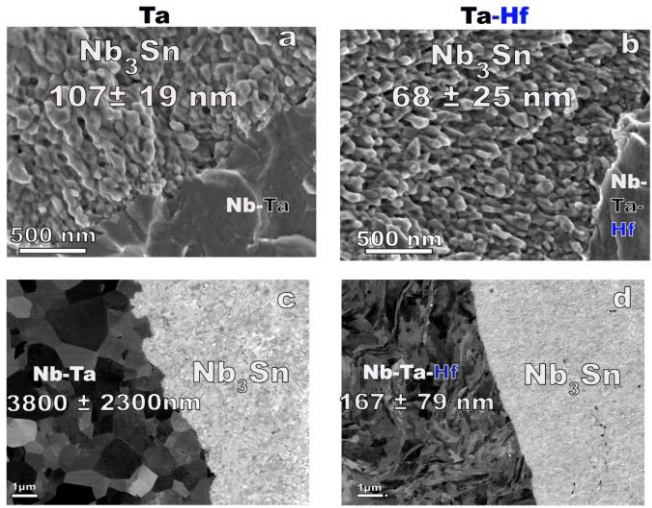

**Figure 1.** Representative fractographs of the $Nb_3Sn$/alloy-rod interface in Ta (a,c) and Ta-Hf (b,d) monofilaments after reaction heat treatment at 550°C/100h + 670°C/100h. Significantly smaller $Nb_3Sn$ grain sizes are obtained in the Ta-Hf sample (b) than in the Ta sample (a). Electron channelling contrast in the corresponding Back-Scattered Electron images shows a recrystallized microstructure in the Nb-Ta alloy rod after reaction heat treatment (c) whereas the unreacted Nb-Ta-Hf alloy rod retains a cold worked microstructure.

## 3 Results

*3.1 Microstructural characterizations*

The in-lens secondary electron (SE) SEM images of fractured wire surfaces of Figure 1(a,b) reveal a typical $Nb_3Sn$ grain structure of the Ta and Ta-Hf wires. The $Nb_3Sn$ grain size of the Ta wire is ~110 nm, comparable to commercial wires. By adding Hf, the A15 grain size noticeably decreases to 68±25 nm and 55±10 nm for Ta-Hf and Ta-Hf-$SnO_2$, respectively. The average grain size for all samples was estimated by the mean lineal intercept method and values are summarized in Table 1.

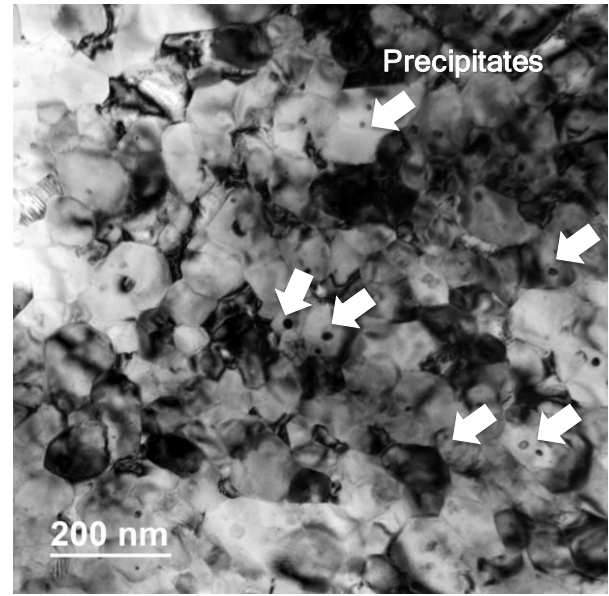

**Figure 2.** Representative bright field TEM image of $Nb_3Sn$ grains close to the alloy-rod interface. White arrows indicate intragranular precipitates in the Ta-Hf sample reacted at 550°C/100h + 670°C/100h.

After the A15 reaction heat treatment, significant Nb alloy rod remains unreacted, leaving a clear microstructural contrast between the alloy rod and the $Nb_3Sn$ as seen in Figure 1(c,d). The channelling contrast in the BSE images highlights evident recrystallization in the Nb4Ta alloy after the A15 reaction heat treatment (Figure 1(c)), while, in strong contrast, showing that the cold worked deformation structure was retained in the Nb4Ta1Hf rod (Figure 1 (d)). The grain size of the unreacted rod was 3.8±2.3 μm in the Nb4Ta but only 0.17±0.08 μm in the Nb4Ta1Hf rod. For comparison, we also measured the grain size of the unreacted alloy rod in the other alloy samples after the A15 reaction heat treatment at 670°C/100h. The grain size of the unreacted alloy rod was 0.29±0.11 μm and 0.310±0.090 μm in Ta-Zr and Ta-Zr-$SnO_2$ samples, respectively. Overall the alloys with $SnO_2$ have slightly smaller grain size than their $SnO_2$-free counterparts.

Figure 2 shows a TEM bright field image of the $Nb_3Sn$ layer close to the interface with the unreacted Ta-Hf alloy rod, which confirms the very small A15 grain sizes (60-70 nm) of the SEM fractograph of Figure 1 (b). In addition to the A15 diffraction grain contrast, there are also dark dots within the $Nb_3Sn$ grains that correspond to nanoparticle precipitation not visible in the fractograph of Figure 1 (b). No visible precipitates were spotted in the TEM images of the unreacted Nb4Ta1Hf rod regions. Quantification and chemical analysis of the precipitates are in progress. Preliminary Electron Energy Loss Spectroscopy (EELS) indicated that the precipitates are Sn deficient without detectable oxygen.

**Table 1**. Summary of Nb$_3$Sn properties in monofilament wires prepared with Ta-Zr/Hf-based ternary alloys after a 670°C/100h heat treatment. Results are compared to a 108/127 Ta-doped RRP conductor heat treated at 666°C.

| Wire Description | | | Grain size | | Electromagnetic properties at 4.2 K | | | | | |
|---|---|---|---|---|---|---|---|---|---|---|
| Sample Name | Alloy | SnO$_2$ | Nb$_3$Sn (nm) | Alloy rod μm | $\mu_0H_{Max}$ T | $F_{pMax}/F_{pMax}^{Nb-Ta}$ | $\mu_0H_{Irr}$ T | Layer $J_c$(12T) A/mm$^2$ | Layer $J_c$(16 T) A/mm$^2$ | Eq.RRP non-Cu $J_c$(16 T), A/mm$^2$ |
| Ta | Nb-Ta | No | 107±19 | 3.8±2.3 | 4.70 | 1 | | 3210±920 | 1250±360 | 750±210 |
| Ta-SnO$_2$ | Nb-Ta | Yes | 100±25 | 2.6±1.9 | 5.10 | 0.65 | | 2240 640 | 1000±290 | 600±170 |
| Ta-Zr | Nb-Ta-Zr | No | 86±27 | 0.29±0.11 | 5.23 | 1.01 | 22.8 | 3550±1010 | 1280±370 | 770±220 |
| Ta-Zr-SnO$_2$ | Nb-Ta-Zr | Yes | 70±33 | 0.31±0.09 | 5.30 | 1.54 | 20.9 | 5020±1430 | 1680±480 | 1010±290 |
| Ta-Hf | Nb-Ta-Hf | No | 68±25 | 0.17±0.08 | 5.81 | 2.35 | 23.6 | 9610±2740 | 3710±1060 | 2230±640 |
| Ta-Hf-SnO$_2$ | Nb-Ta-Hf | Yes | 55±10 | 0.13±0.06 | 5.47 | 2.25 | 23.1 | 8520±2430 | 3090±880 | 1860±530 |
| Ta-RRP | Nb-Ta | - | | | 4.60 | | 23.2 | | | |

Since the Nb$_3$Sn layers formed by Sn diffusion develop compositional and/or microstructural gradients away from the Sn source, we also examined the Nb$_3$Sn layer close to the Cu-Sn region, confirming that the Sn-rich Nb$_6$Sn$_5$ phase was also produced in this rod-in-powder-in-tube design. Figure 3 is a TEM image of the interface between the outer Nb$_6$Sn$_5$ and the inner Nb$_3$Sn layer (the unreacted Nb4Ta1Hf rod lies inside these layers). An interesting observation is the presence of very small equiaxed Nb$_6$Sn$_5$ grains of less than 100 nm diameter. This is one of the first observations of sub-micron Nb$_6$Sn$_5$, whose grain size is normally several microns in commercial PIT conductors [13].

*3.2 Magnetization characterizations*

Magnetization measurements were performed in the 1.3-14 K temperature range on all the monofilament wires and, for comparison, a Ta-doped RRP® conductor which was similarly heat treated (666°C/50h). The 35 T magnet allowed us to evaluate the entire hysteresis loops so as to accurately estimate the temperature dependence of $H_{Irr}$. The layer $J_c$ and $F_p$ were estimated using the Bean model combined with the cross-sectional dimensions determined by SEM image analysis. The A15 Thickness and area fraction was identified based on BSE contrast differences. Because of the simple wire design and some non-uniform deformation of the Nb alloy rods during wire drawing, a noticeable variation in the residual rod cross-section was found within each wire, affecting the calculation of Nb$_3$Sn cross-sectional area. To estimate the layer $J_c$, we analysed 10 different wire cross-sections to determine the average rod and Nb$_3$Sn cross-sectional areas and their respective standard deviations.

The deduced layer $F_p$ and $J_c$ values are plotted in Fig. 4 and in the inset of Fig. 6. For the standard Nb4Ta alloy (without SnO$_2$) $F_p$ peaks at $H_{Max}$ ~ 4.7 T, close to the value for Ta-doped RRP® wire (see Table 1). Adding SnO$_2$ to the Ta-doped wire shifts $H_{Max}$ above 5 T but clearly supresses $F_{pMax}$ by 35%. The Nb4Ta1Zr alloy shifts $H_{Max}$ to ~ 5.2-5.3 T with $F_{pMax}$ for the wire without SnO$_2$ being comparable to that of the standard Nb4Ta alloy, whereas the wire with SnO$_2$ addition has a more than 50% larger $F_{pMax}$. These two curves however cross at high field due to some $H_{Irr}$ degradation when SnO$_2$ is present. The Nb4Ta1Hf alloy is clearly the most effective in changing the pinning force, having $H_{Max}$ peaking at ~5.5 and ~5.8 T for the wires with and without $SnO_2$, respectively. $F_{pMax}$ values in these cases are comparable and both more than 2.2 times larger than for the standard Nb4Ta alloy.

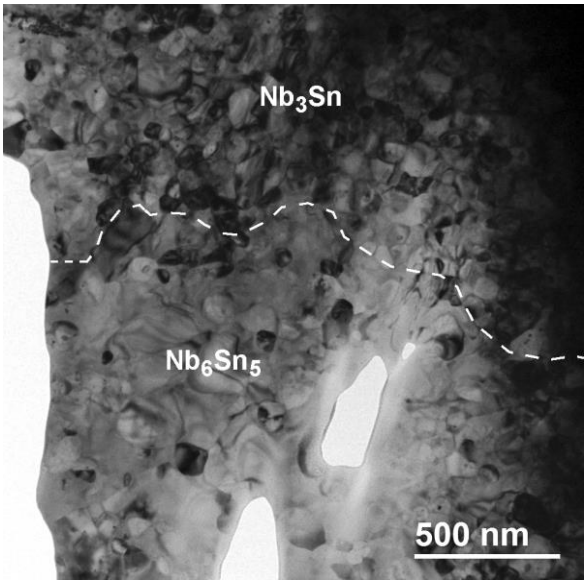

**Figure 3.** Representative TEM bright field image of the Nb$_3$Sn-Nb$_6$Sn$_5$ interface in the Ta-Hf sample reacted at 550°C/100h + 670°C/100h. Grain sizes of both Nb$_6$Sn$_5$ and Nb$_3$Sn are below 100 nm.

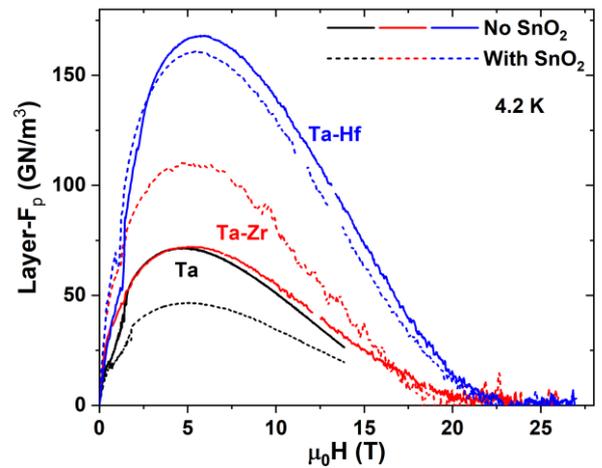

**Figure 4.** Pinning force density of the A15 layer as a function of magnetic field showing significant enhancement for the Nb-Ta-Zr and Nb-Ta-Hf conductors.

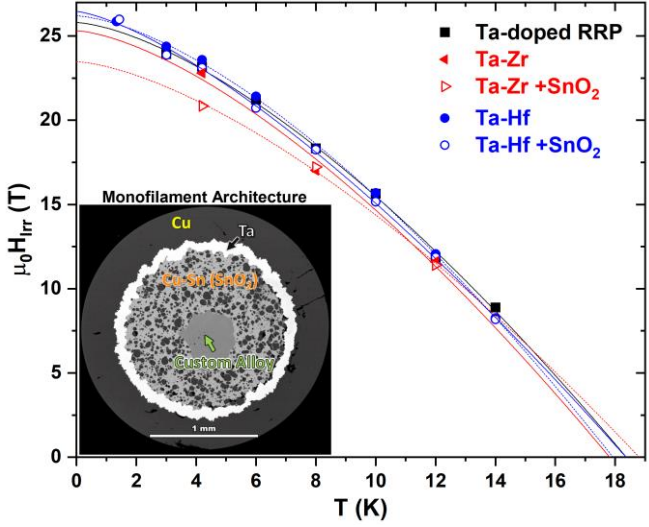

**Figure 5.** Irreversibility field versus temperature determined from the magnetization hysteresis of the various monofilaments compared with a Ta-doped RRP conductor. $H_{Irr}$ of the Ta-Hf doped monofilament is the highest and both Zr and Hf monofilaments with $SnO_2$ show degraded $H_{Irr}$. The inset shows an unreacted monofilament cross-section.

The temperature dependence of $H_{Irr}$ is shown in Fig. 5. Comparing the 4.2 K values, we found that Ta-Zr, and Ta-Hf without $SnO_2$ have $H_{Irr}$ values of 22.8 and 23.6 T, respectively, similar to $H_{Irr}$ of 23.2 T obtained for the RRP conductor. The $SnO_2$ addition to the Ta-Zr sample supresses $H_{Irr}$(4.2 K) to 20.9 T, whereas in the Ta-Hf case $H_{Irr}$ had a smaller drop from 23.6 to 23.1 T. Overall, oxygen appears to degrade both Hf and Zr-doped conductors but the effect is larger for the Nb4Ta1Zr alloy.

Figure 6 and Table 1 summarize the layer $J_c$(4.2 K) at 12 and 16 T of our monofilaments and compare them to commercial conductors: the grey shaded areas represent the range of variation for high-$J_c$ Ta-doped RRP® from ref. 31, while the orange dashed lines are for the best Ti-doped RRP® after the optimized Cu-Sn mixing heat treatment of ref. 8. Our Nb4Ta wire has a layer $J_c$ of 3210 ± 920 A/mm² at 12 T and 1250 ± 360 A/mm² at 16 T, values typical of those found for Ta-doped RRP® conductors, which range between 3190-5250 A/mm² at 12 T and 1400-1880 A/mm² at 16 T. The Ti-doped RRP® conductor after optimized heat treatment has markedly better 16 T properties, with a layer $J_c$ > 2000 A/mm². Oxygen addition to the Nb4Ta monofilament reduces the layer $J_c$ below the oxygen-free Nb4Ta values (2200 ± 640 A/mm² at 12 T and 1000 ± 290 A/mm² at 16 T). Additions of 1Zr to Nb4Ta improve the layer $J_c$ of the conductor to 3550 ± 1010 A/mm² (12 T) and 1280 ±

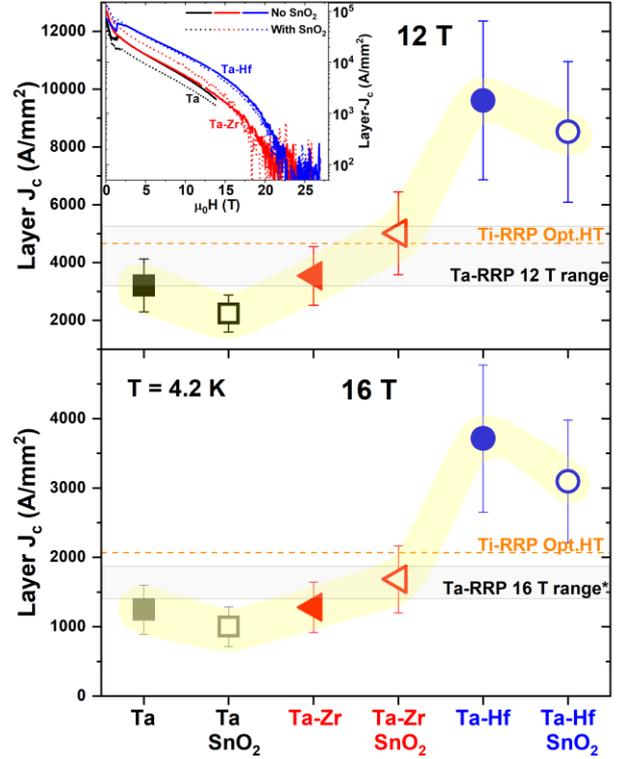

**Figure 6.** Layer $J_c$(4.2 K) of $Nb_3Sn$ in monofilament wires at 12 and 16 T (4.2 K) showing significantly better performance in the Ta-Hf based conductors with respect to the standard Ta and Ta-Zr based wires by about a factor of three. The grey shaded areas represent the range of variation for high-$J_c$ Ta-doped RRP® [31]; the orange dash lines are for the best Ti-doped RRP® wire after optimized heat treatment [8]. The 16 T data for the Ta and Ta-$SnO_2$ samples and the Ta-RRP conductors were estimated from 14 T measurements.

370 A/mm² (16 T), and oxygen addition further improves the layer $J_c$ to 5020 A/mm² at 12 T and 1680 A/mm² at 16 T. However, the most interesting results are obtained in the Nb4Ta1Hf monofilaments where the layer $J_c$ without $SnO_2$ is well above all other conductors, exceeding 9600 A/mm² at 12 T and 3710 ± 1060 A/mm² at 16 T. Unlike its Nb4Ta1Zr counterpart, the Hf conductor shows a slight layer-$J_c$ decrease with oxygen addition. However, the values are still very high (8520 A/mm² at 12 T and 3090 A/mm² at 16 T).

## 4. Discussion

Despite the observation of a upward shift in $F_{pMax}$ towards higher field, earlier Nb1Zr PIT conductors showed degraded $H_{Irr}$ of ~20 T or less, lower than binary $Nb_3Sn$, thus depressing the high field $J_c$ properties significantly below standard Ta or Ti-doped conductors [8,14]. A new result of this work is that $H_{Irr}$ suppression is no longer an issue for conductors fabricated with Nb4Ta alloy with 1Zr or 1Hf additions. The A15 $H_{Irr}$(4.2K) now reaches 23.6 T in the O-free Hf alloy without any degradation compared to commercial wires made with Nb4Ta and Nb-Ti. The beneficial effects of Zr added to Nb4Ta in a PIT wire have also been recently reported by Xu et al. [32]. However, in this comparison of properties with and without SnO$_2$, we note that we do still see a small $H_{Irr}$ degradation in the presence of O, the effect being significantly worse in the Zr alloy. Although showing an improved $H_{Max}$ of 5.3 T, the Nb4Ta1Zr alloy with SnO$_2$ has $H_{Irr}$(4.2 K) of only 20.85 T, ~2 T lower than Nb4Ta1Zr without SnO$_2$.

Perhaps the most surprising result of the present study is that the Nb$_3$Sn grain size can be reduced to much less than 100 nm without SnO$_2$ and its internal oxidation potential when Nb4Ta1Hf alloy is used, perhaps because some nano-precipitates form in the A15 layer even without internal oxidation by SnO$_2$. The explanation may lie in another unexpected observation, namely that neither the Nb4Ta1Zr nor the Nb4Ta1Hf alloys recrystallized during the A15 reaction heat treatment, quite unlike the case with Nb4Ta (or pure Nb) rods. As seen in Figure 1 and Table 1, the grain structure of the Nb4Ta1Zr and Nb4Ta1Hf rods after A15 reaction heat treatment is one order of magnitude smaller than in the recrystallized Nb4Ta alloy. Since penetration of Sn into the alloy must occur preferentially by grain boundary diffusion [33], the much denser diffusion paths available to the unrecrystallized ternary alloys must contribute significantly to denser A15 nucleation and a finer A15 grain size in the growing reaction layer. Indeed we propose that a high A15 nucleation frequency in the cold-worked alloy nanostructure with its high GB density is the dominant factor to significantly reduce the Nb$_3$Sn grain size in our Ta-Zr and Ta-Hf samples. Our observation of 100 nm grains in the intermediate Nb$_6$Sn$_5$ phase which forms between the Sn source and the A15 layer is also consistent with this same Sn diffusion mechanism along the Nb alloy GBs. SnO$_2$ addition does appear to produce a small decrease in A15 grain size (from 68 to 55 nm), perhaps by providing some precipitation that retards A15 grain growth. In this respect the 1Hf alloy appears better than the 1Zr alloy precisely because it can strongly enhance vortex pinning without need for SnO$_2$ and without the $H_{Irr}$ degradation that O brings in these wires.

In considering the practical implications of these experiments, we note that the highest non-Cu $J_c$(4.2 K, 16 T) obtained by optimized heat treatment of state-of-the-art commercial Ti-doped RP® conductor is so far only ~1300 A/mm$^2$. This new record for an RRP conductor emerged from a detailed study of a wide range of RRP billets made available through the US Magnet Development Program and Oxford Superconducting for study by Sanabria et al.[8], most of which were of lower $J_c$ due to the inherent distribution of properties seen in large production runs. The present results have a particular excitement that the highest $J_c$ was obtained in the Nb4Ta1Hf alloy without SnO$_2$, suggesting that conventional RRP® or PIT architectures are possible with minimum perturbation to present billet designs. The RRP® version might thus create <100 nm Nb$_3$Sn grains with a layer $J_c$(4.2 K, 16 T) of ~3700 A/mm$^2$. Considering that the Nb$_3$Sn layer occupies about 60% of the non-Cu area in RRP® wires, this is equivalent to a non-Cu $J_c$(4.2 K, 16 T) of ~2200 A/mm$^2$, well above the FCC requirement of 1500 A/mm$^2$. Moreover, the Nb1Ta4Hf alloy could also be beneficial for PIT conductors presently made with Nb4Ta tubes. Detailed studies of the A15 reaction in present PIT conductors by Segal et al. show that $J_c$ is in fact limited by the fact that about one quarter of the ~55% A15 (formed by decomposition of large grain Nb$_6$Sn$_5$) is >1 µm in size [12,13,34]. This large grain A15 does not contribute to current transport, thus degrading the overall $J_c$ of PIT conductors with respect to RRP® conductors. Segal's studies showed that the large grain A15 could be partially suppressed in favour of fine-grain A15 at higher reaction temperatures. $J_c$ did not benefit however due to the larger A15 grain size of higher temperature reactions, presumably associated with grain growth in the recrystallized Nb4Ta. Given the ability of Nb4Ta1Hf to avoid recrystallization, it may be that the inverted higher-lower temperature A15 reactions studied by Segal now become feasible.

In summary we have shown that 1Hf additions to Nb4Ta show many benefits and suggest straightforward routes to much higher $J_c$ Nb$_3$Sn conductors made by all of the present commercial routes, bronze, internal Sn and other variants. A potential concern may be that both Zr and Hf additions to Nb4Ta degrade alloy drawability, but, so far, after a true strain of 8, no problems of work hardening and filament breakage have been encountered. Further studies to higher strains are in progress.

## 5. Conclusions

We have demonstrated that adding 1Hf or 1Zr to Nb4Ta greatly reduces the $H_{Irr}$ suppression encountered in Nb$_3$Sn wires made from Nb1Zr wires previously examined at OSU and SupraMagnetics/FSU [28-30]. The highest $H_{Irr}$ and the best vortex pinning performance were obtained with a Nb4Ta1Hf alloy without SnO$_2$, which reached a maximum pinning force more than twice that obtained with standard Nb4Ta alloy and with slightly better $H_{Irr}$. Initial evaluations of the layer $J_c$ lead to ~3700 ± 1100 A/mm$^2$ at 16 T, 4.2 K corresponding to a non-Cu $J_c$ of ~2200 ± 600 A/mm$^2$ in an RRP architecture. These results indicate that the quickest pathway to a high $J_c$ for FCC conductor may be to avoid internal oxidation and to just use the A15 grain refinement properties of Hf or similar additions to Nb4Ta in either internal-tin or PIT conductors.

**Acknowledgements**

The authors would like to thank to J. Parrell (Bruker OST, LLC) for providing Nb7.5 Ta slabs used for the special alloys melted here and Eun Sang Choi (NHMFL) for his support during high-field VSM characterization. The authors would like to acknowledge the TEM sample preparation work of undergraduate students T. Shelby and J. Boyd (NHMFL-ASC). Discussions with Chris Segal (now at CERN) are gratefully acknowledged. This work is funded by the U.S. Department of Energy, Office of Science, and Office of High Energy Physics under Award Number DE-SC0012083, and performed under the purview of the US-Magnet Development Program. Discussions of parallel work being performed by Xingchen Xu (Fermilab) are greatly acknowledged. This work was performed at the National High Magnetic Field Laboratory, which is supported by National Science Foundation Cooperative Agreements NSF-DMR-1157490 and DMR-1644779 and by the State of Florida.